\documentclass[aps,prb,twocolumn,superscriptaddress,showpacs,floatfix]{revtex4}
\usepackage{amsmath,amssymb}
\usepackage{graphicx}
\newcommand{\bfj}{{\mathbf{j}}}
\newcommand{\bfk}{{\mathbf{k}}}
\newcommand{\bfn}{{\mathbf{n}}}
\newcommand{\bfp}{{\mathbf{p}}}
\newcommand{\bfr}{{\mathbf{r}}}
\newcommand{\bfv}{{\mathbf{v}}}
\newcommand{\bfx}{{\mathbf{x}}}
\newcommand{\bfy}{{\mathbf{y}}}
\newcommand{\bfz}{{\mathbf{z}}}
\newcommand{\varH}{{\mathcal{H}}}
\newcommand{\bfnabla}{{\boldsymbol{\nabla}}}
\newcommand{\bfsigma}{{\boldsymbol{\sigma}}}
\newcommand{\re}{\,\mathrm{Re}}

\newcommand{\ux}        {{\hat\bfx}}
\newcommand{\uy}        {{\hat\bfy}}
\newcommand{\uz}        {{\hat\bfz}}
\newcommand{\nm}        {\mathrm{nm}}
\newcommand{\umeter}    {\mu\mathrm{m}}

\newcommand{\angstrom}  {\mathrm{\AA}}
\newcommand{\eV}        {\mathrm{eV}}
\newcommand{\meV}       {\mathrm{meV}}

\newcommand{\figref}[1]{Fig.~\ref{#1}}
\newcommand{\Figref}[1]{Figure~\ref{#1}}

\begin{document}
\title{Ballistic spin currents in mesoscopic metal/In(Ga)As/metal junctions}%
\author{Minchul Lee}%
\affiliation{Department of Physics, Korea University, Seoul 136-701,
  Korea}%
\affiliation{Department of Physics and Astronomy, University of Basel,
  CH-4056 Basel, Switzerland}
\author{Mahn-Soo Choi}%
\affiliation{Department of Physics, Korea University, Seoul 136-701,
  Korea}%

\begin{abstract}
We investigate the ballistic spin transport through a two-dimensional
mesoscopic metal/semiconductor/metal double junctions in the presence of
spin-orbit interactions.  It is shown that \emph{real} longitudinal
and/or transverse spin currents can flow in the presence of the Rashba
and Dresselhaus terms.
\end{abstract}

\pacs{73.63.-b, 72.10.-d, 71.70.Ej}
\maketitle

\section{Introduction}

Since the advent of ``spintronics'' to utilize electron's spin rather
than its charge for information processing and
storage,\cite{Spintronics} there has been growing interest in generating
spin currents in diverse ways.\cite{Hammar99,Ohno99,ESHE,ISHE1,ISHE2}
Though injecting spin-polarized carriers electrically remains a
challenge,\cite{Hammar99} various kinds of all-semiconductor devices
using ferromagnetic semiconductor heterostructure\cite{Ohno99} or
spin-orbit (SO) interactions\cite{ESHE} have been proposed. The
so-called \emph{extrinsic spin Hall effect} due to SO dependent
scattering from magnetic impurities manifests a spin current because of
a transverse spin imbalance generated from a charge current circulating
in a paramagnetic metal.\cite{ESHE} In the weak impurity scattering
limit but with substantial SO couplings, on the other hand, it was
suggested that the \emph{intrinsic spin Hall effect} gives rise to a
dissipationless spin current perpendicular to the external electric
field.\cite{ISHE1,ISHE2} Moreover, the spin Hall conductance has a
universal value.  However, it was demonstrated \cite{Rashba03} that the
dissipationless (unreal) spin current does not vanish even in
thermodynamic equilibrium in the absence of external fields, putting the
interpretation of intrinsic spin Hall effect in a controversy.

\begin{figure}
\centering%
\includegraphics*[width=75mm]{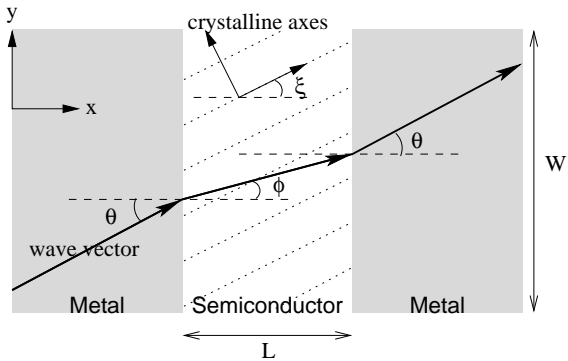}
\caption{A schematic of the system.}
\label{bsc::fig:1}
\end{figure}

In this paper we study the ballistic spin transport through a mesoscopic
double-junction system consisting of a semiconductor stripe sandwiched
by two normal metal leads; see Fig.~\ref{bsc::fig:1}.  We use the
coherent scattering theory and show that in the presence of SO
couplings, both longitudinal and transverse spin currents can flow in
the semiconductor.  It is stressed that these currents are real; see
Ref.~\onlinecite{Rashba03}.

\section{Model and Scattering Theory}

We consider a two-dimensional electron system (2DES) of semiconductor
(S) between two normal (N) metal leads.  We choose such a coordinate
system that $x$-axis ($y$-axis) is perpendicular (parallel) to the N/S
interfaces and $z$-axis is perpendicular to the 2D plane; see
Fig.~\ref{bsc::fig:1}.  The length (width) of the semiconductor is $L$
($W$); we will consider the limit $W\to\infty$.  Within the
effective-mass approximation,\cite{EffectiveMassApproximation} the
Hamiltonian reads as
\begin{equation}
\varH = -\frac{\hbar^2}{2}\bfnabla\cdot\frac{1}{m(x)}\bfnabla
+ V(x,y) + \varH_R(x) + \varH_D(x).
\end{equation}
The position-dependent effective mass $m(x)$ has values of $m_e$ and
$m_e^*\equiv\epsilon_m m_e$ in the normal metals and the semiconductor
($-L/2\!<\!x\!<\!L/2$), respectively. The confinement potential has a potential
barrier of height $V_0$ inside the semiconductor:
\begin{equation}
V(x,y) = V_0\left[\Theta(x+L/2)-\Theta(x-L/2)\right] + V(y),
\end{equation}
where $\Theta(x)$ is the Heaviside step function and $V(y)$ accounts for
the finite width $W$.  The potential barrier height $V_0$ is lower than
the Fermi energy $E_F$ in the normal metals so that $E_F^* \equiv E_F -
V_0 >0$.
The Rashba \cite{Rashba84} and Dresselhaus \cite{Dresselhaus55} SO
coupling terms are given by
\begin{equation}
\label{eq:SOH}
\varH_R = \frac{\alpha}{\hbar} (\sigma_xp_y{-}\sigma_yp_x)\ \ \text{and}\ \
\varH_D = \frac{\beta}{\hbar} (\sigma_yp_y{-}\sigma_xp_x),
\end{equation}
respectively, inside the semiconductors while they vanish in the normal
metal sides.  In Eq.~\eqref{eq:SOH},
$\bfsigma=(\sigma_x,\sigma_y,\sigma_z)$ are the Pauli matrices.

The Rashba term $\varH_R$ arises when the confining potential of the
quantum well lacks the inversion symmetry, while the Dresselhaus term
$\varH_D$ is due to the bulk inversion asymmetry.  In some semiconductor
heterostructures (e.g., InAs quantum wells) $\varH_R$
dominates\cite{Das90a}, and in others (e.g., GaAs quantum wells)
$\varH_D$ is comparable to (or even dominant over)
$\varH_R$\cite{Lommer88a}.  The coupling constants may range around
$\alpha\sim 0.1\,\eV{\cdot}\angstrom$ and $\beta\sim
0.09\,\eV{\cdot}\angstrom$, respectively, depending on the
structure and material.\cite{Schliemann03a}

\begin{figure}
\centering%
\includegraphics[draft=false,width=4cm]{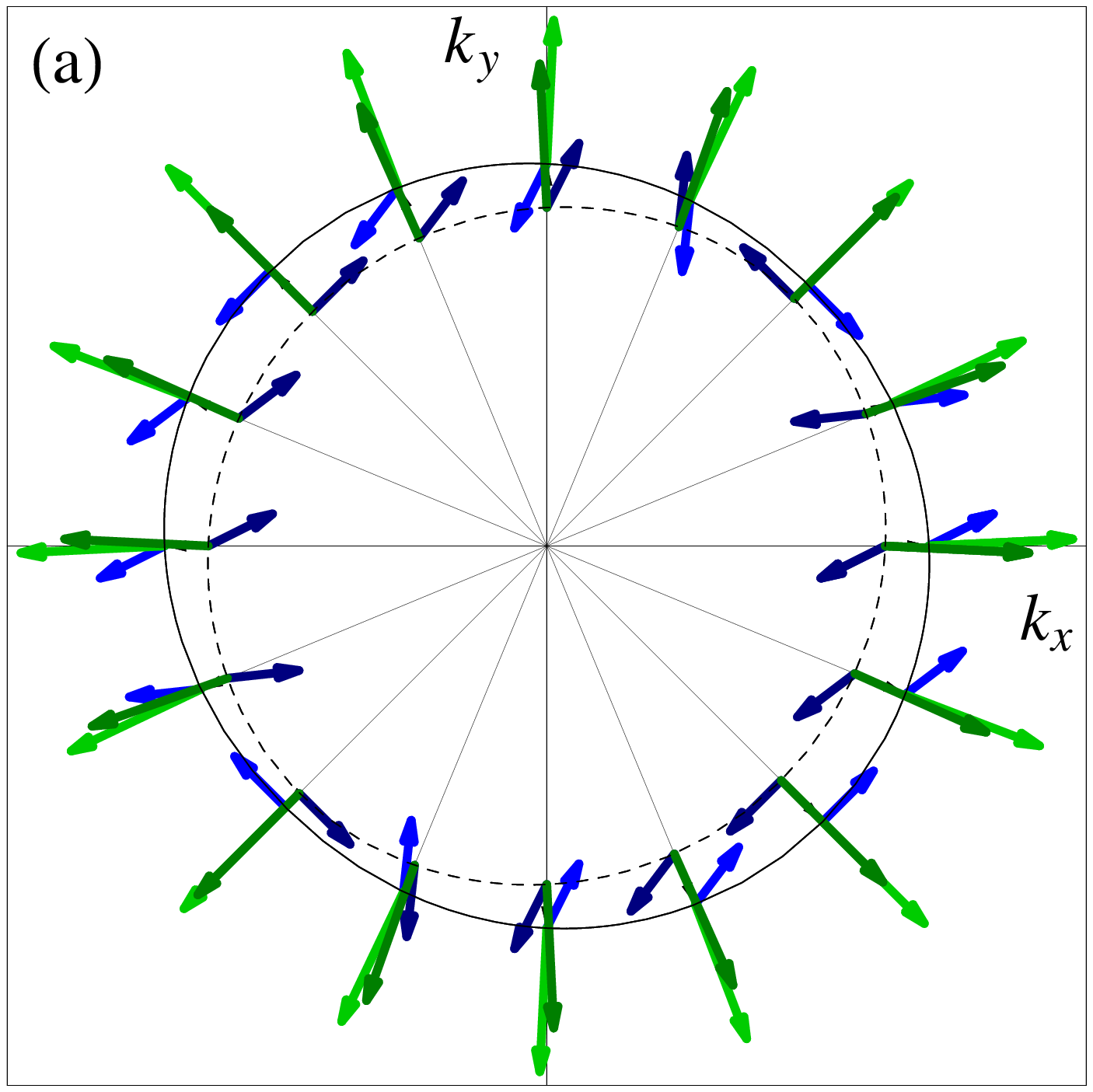}
\includegraphics[draft=false,width=4cm]{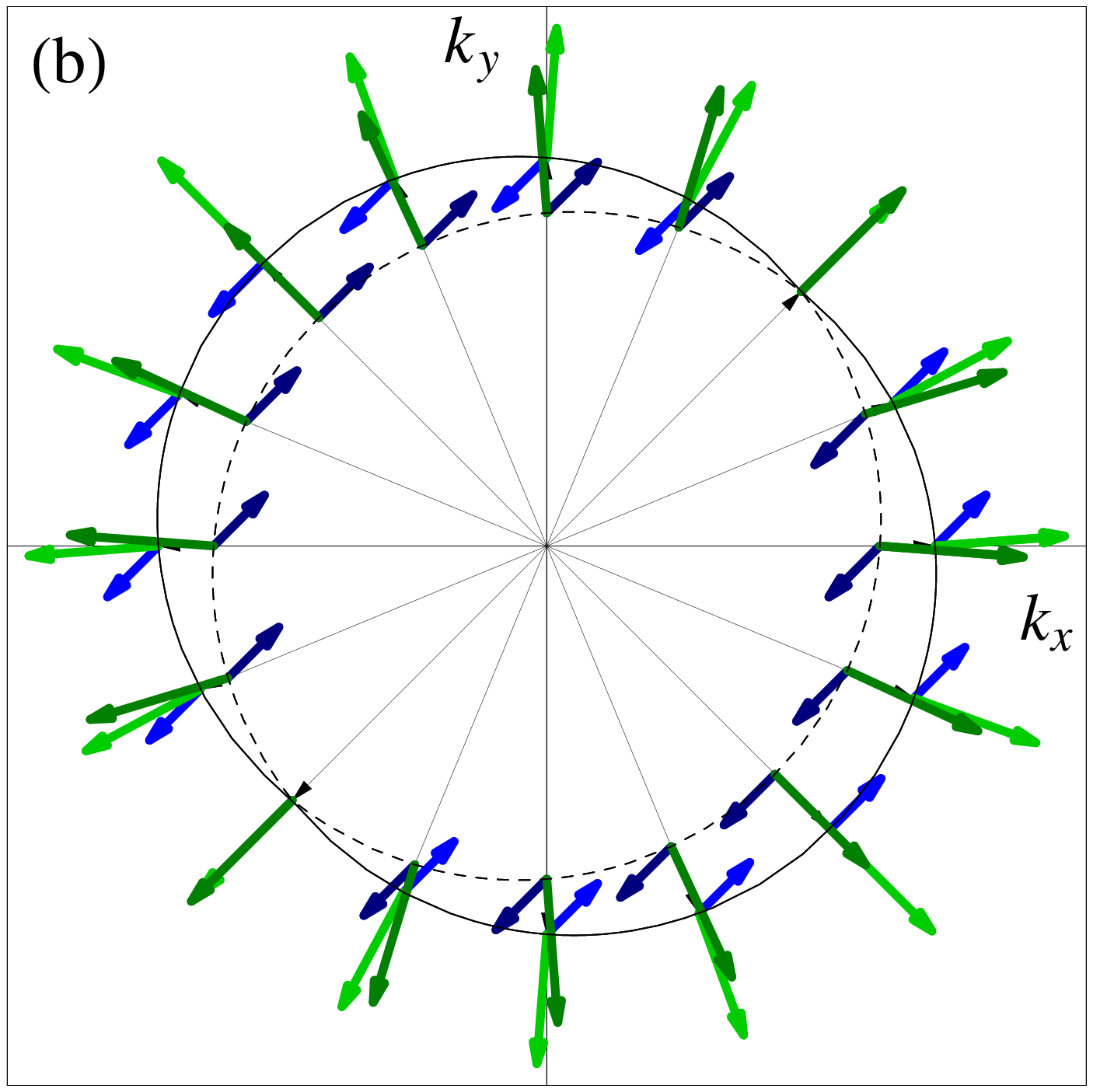}
\caption{Fermi contours (solid/dotted lines for $\mu=\pm$), and on each
  of them, the wave vectors $\bfk$ (black solid arrows), the group
  velocities $\bfv^\pm_\bfk$ (dark/light green arrows), and the spin
  orientations $\pm\hat\bfn_\bfk$ (dark/light blue arrows) of the
  eigenstates.  (a) $\alpha=0.5\beta$ and (b) $\alpha=\beta$ with $E_F^*
  = 14\,\meV$, $\beta=0.1\,\eV{\cdot}\angstrom$, and
  $\epsilon_m=0.063$.}
\label{bsc::fig:2}
\end{figure}

Inside the semiconductor, electrons feel a fictitious, in-plane magnetic
field in the direction
\begin{equation}
\label{bsc::eq:n}
\hat\bfn_\bfk = \ux\cos\varphi_\bfk + \uy\sin\varphi_\bfk \,,
\end{equation}
where
\begin{math}
\varphi_\bfk
= \arg[(\beta k_x - \alpha k_y) + i(\alpha k_x - \beta k_y)]
\end{math}.
Accordingly, the eigenstates with spin parallel ($\mu=+$) and
anti-parallel ($\mu=-$) to $\hat\bfn_\bfk$ for a given wave vector
\begin{math}
\bfk = k(\ux\cos\phi + \uy\sin\phi)
\end{math}
are written in the spinor form
\begin{equation}
\label{eq:eigenstate}
\Psi^\mu_\bfk(\bfr) = \frac{e^{i\bfk\cdot\bfr}}{\sqrt{2}}
\begin{bmatrix}
\mu e^{-i\varphi_\bfk/2} \\ e^{+i\varphi_\bfk/2}
\end{bmatrix} \,.
\end{equation}
The corresponding energies are
\begin{equation}
\label{eq:dispersion}
E_\mu(\bfk) = \frac{\hbar^2}{2m_e^*}[k^2-2\mu k_{so}(\phi)k] \,,
\end{equation}
where
\begin{math}
k_{so}(\phi) \equiv (m_e^*/\hbar^2) %
\sqrt{\alpha^2+\beta^2 - 2\alpha\beta\sin2\phi}
\end{math}.
From the continuity equation for the charge density, one can get the
expression for the charge current associated with a given wave function
$\Psi(\bfr)$\cite{Matsuyama02}
\begin{equation}
\bfj_c = e\re\left[\Psi^\dag(\bfr)\bfv\Psi(\bfr)\right] \,,
\end{equation}
where $\bfv$ is the velocity operator defined by
\begin{equation}
\bfv = \frac{\bfp}{m_e^*}
- \frac{\alpha}{\hbar} (\sigma_y\ux-\sigma_x\uy)
- \frac{\beta}{\hbar}(\sigma_x\ux-\sigma_y\uy) \,.
\end{equation}
In the same manner, we define the spin current\cite{Rashba03}
\begin{equation}
\label{bsc::eq:Js}
\bfj_s(\hat\bfn) = \frac{\hbar}{2} \Psi^\dag(\bfr)\frac{\bfv
  (\hat\bfn\cdot\bfsigma) + (\hat\bfn\cdot\bfsigma)\bfv}{2}\Psi(\bfr)
\end{equation}
according to the continuity equation
\begin{equation}
\label{bsc::eq:SCE}
\partial_t Q_s + \bfnabla\cdot\bfj_s = S_s
\end{equation}
for the spin density (with respect to the spin direction $\hat\bfn$)
\begin{equation}
Q_s(\hat\bfn) \equiv
\frac{\hbar}{2}\left[\Psi^\dag(\bfr) (\hat\bfn\cdot\bfsigma)
  \Psi(\bfr)\right]
\end{equation}
and the spin source
\begin{equation}
\label{bsc::eq:spin-source}
S_s(\hat\bfn) =\frac{\hbar}{2}
\re\left[\Psi^\dag(\bfr)\frac{i}{\hbar}
  \left[\varH,\hat\bfn\cdot\bfsigma\right]\Psi(\bfr)\right] \,.
\end{equation}
The appearance of the spin source term in Eq.~\eqref{bsc::eq:SCE} is not
surprising because the spin-orbit couplings break spin conservation
inside the semiconductor.

Before going further, it will be useful to understand the origin of the
spin current in physical terms.
As illustrated in Fig.~\ref{bsc::fig:2}, for $\alpha,\beta\neq 0$
the Fermi contours,
\begin{equation}
\label{bsc::eq:contour}
k_F^\mu(\phi) = \mu k_{so}(\phi) + \sqrt{k_{so}^2(\phi)+k_F^{*2}}
\end{equation}
with $k_F^*\equiv\sqrt{2m^*E_F^*}/\hbar$, are no longer isotropic, and
the group velocities
\begin{math}
\bfv_\bfk^\mu = \Psi_\bfk^{\mu\dagger}\bfv\Psi_\bfk^\mu
\end{math}
of the eigenstates in Eq.~\eqref{eq:eigenstate} are not parallel to the
wave vector $\bfk$.~\cite{ScatteringFormalism,Matsuyama02,Schliemann03b}
Nevertheless, Eq.~\eqref{bsc::eq:contour} reveals an important symmetry
property of the group velocities:
\begin{math}
\left|\bfv^+_{\bfk_F^+}\right| =
\left|\bfv^-_{\bfk_F^-}\right|
\end{math}.
It means that the two eigenstates with opposite spin orientations make
the same contributions to the charge transport along the $\hat\bfk$
direction (and opposite contributions along the the perpendicular
direction).  On the contrary, for the spin transport with $\hat\bfn =
\hat\bfn_{\hat\bfk}$, the inverse occurs: two eigenstates make opposite
contributions along the $\hat\bfk$ direction and same ones in the
perpendicular direction.  This implies the possibility of the
observation of the net spin current flowing perpendicular to the charge
current.
Particularly interesting are the cases of $\alpha=\pm\beta$, where all
the spin orientations, $\pm\hat\bfn_\bfk$, for different wave vectors are
parallel or anti-parallel to each other ($\varphi_\bfk=\pi/4$); see
\figref{bsc::fig:2}~(b). It results from conservation of
$(\sigma_x\pm\sigma_y)/\sqrt{2}$, and the spin state become independent
of the wave vector.\cite{Schliemann03a,Schliemann03b}

Now we study charge and spin transport in N/S/N double junction
structures.  Coherent scattering formalism at the N/S interfaces has
been thoroughly developed in the previous
studies,\cite{ScatteringFormalism,Matsuyama02} considering the Rashba SO
effect and appropriate boundary conditions requiring the conservation of
probability current normal to the interface. It is straightforward to
extend the scattering theory to incorporate the Dresselhaus effect. We
have used the transfer-matrix formalism to calculate the conductance
through and inside the semiconductor: for details refer to
Refs.~\onlinecite{Matsuyama02,Usaj04}.

We will consider electrons incident from the left lead and reflected
from the junction interfaces or transmitted through them to the right
lead.  The wave vector of the incident electron is at angle $\theta$
with respect to the normal to the interface; see Fig.~\ref{bsc::fig:1}.
Contrary to the Rashba effect, the Dresselhaus effect is not invariant
under the rotation, leading to anisotropic
transport.\cite{Schliemann03b} Hence the relative orientation, $\xi$, of
the crystal symmetry axes and the interface (see Fig.~\ref{bsc::fig:1})
affects especially the spin current significantly.
Below we will calculate the charge conductance $G_\nu^{(c)}(\theta)$
($\nu=x,y$) in the $\nu$-direction for a definite incident angle
$\theta$ as well as the angle-averaged quantity
\begin{math}
G_\nu^{(c)}
= \int_{-\pi/2}^{\pi/2}d\theta\;G_\nu^{(c)}(\theta)
\end{math}.
Also calculated are the (analogously defined) spin conductances
$G_\nu^{(s,\hat\bfn)}(\theta)$ and $G_\nu^{(s,\hat\bfn)}$ polarized in
the direction $\hat\bfn$.
The typical values for the parameters we will use below are
$E_F=4.2\,\eV$, $\epsilon_m=0.063$, $\beta=0.1\,\eV{\cdot}\angstrom$,
$L=200\,\nm$, and $W=1\,\umeter$.  $\alpha$ ranges from $-2\beta$ to
$+2\beta$, and $E_F^*$ from $0$ to $20\,\meV$.
We assume sufficiently low temperatures ($k_BT\ll E_F^*$).

\section{Normal Incidence}

\begin{figure}
\centering%
\includegraphics*[width=8cm]{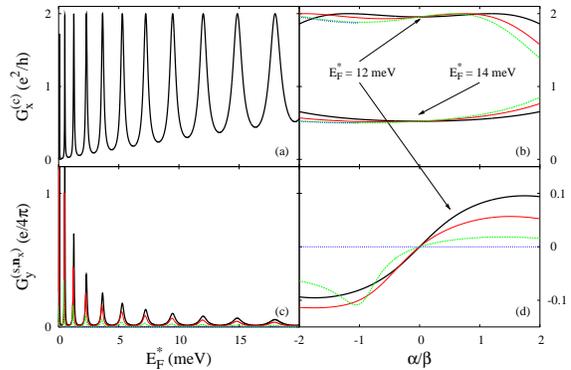}
\caption{The charge conductance $G_x^{(c)}(\theta\!=\!0)$ [(a) and
  (b)] and the spin conductance $G_y^{(s,\hat\bfn_\ux)}(\theta\!=\!0)$
  [(c) and (d)] for normal incidence as functions of $E_F^*$ [(a) and
  (c)] and $\alpha/\beta$ [(b) and (d)]. $\alpha/\beta=0.5$ in (a) and
  (b), and $E_F^*=12$ and $14\,\meV$ in (b) and (d). $\xi$ has been
  chosen to be 0 (black line), $\frac{\pi}{10}$ (red solid line),
  $\frac{\pi}{5}$ (green dashed line), and $\frac{\pi}{4}$ (blue dotted
  line).  Notice that in (a) curves for different $\xi$'s overlap almost
  completely.}
\label{bsc::fig:3}
\end{figure}

Owing to the symmetry
\begin{math}
\left|\bfv_{\bfk_F^+}^+\right| = \left|\bfv_{\bfk_F^-}^-\right|
\end{math}
[see the discussion below Eq.~\eqref{bsc::eq:spin-source}],
for normal incidence ($\theta\!=\!0$) the charge current is completely
longitudinal; i.e, $G^{(c)}_y(\theta\!=\!0) = 0$.
For a single transverse mode, we obtain
the longitudinal charge conductance
\begin{equation}
\label{bsc::eq:Gcx:1}
G^{(c)}_x(\theta\!=\!0)
= \frac{e^2}{h}%
\frac{32\kappa^2}{|(1{+}\kappa)^2 {-}
  (1{-}\kappa)^2 e^{2i\Delta kL}|^2} \,,
\end{equation}
where $\Delta k \equiv \sqrt{k_{so}^2(-\xi)+k_F^{*2}}$,
$\kappa\equiv\Delta k/\epsilon_m k_F$,
$k_F\equiv\sqrt{2m_eE_F}/\hbar$.
Moreover, the spin current has only transverse component and is
polarized entirely in the $xy$-plane; i.e.,
$G^{(s,\hat\bfn)}_x(\theta\!=\!0) = 0$ for any $\hat\bfn$ and
$G^{(s,\hat\bfz)}_y(\theta\!=\!0) = 0$.
The $\hat\bfn_\ux$-polarized spin conductance
$G_y^{(s,\hat\bfn_\ux)}(\theta\!=\!0)$ is given by
\begin{multline}
\label{eq:gsyn}
G_y^{(s,\hat\bfn_\ux)}(\theta\!=\!0) =
\frac{e}{4\pi}\frac{L}{W}
\frac{32(m_e^{*2}/\hbar^4)\alpha\beta\cos2\xi}{\epsilon_m k_F
  k_{so}(-\xi)} \\\mbox{}\times %
\frac{(1{+}\kappa^2) -
  (1{-}\kappa^2)\frac{\sin 2\Delta kL}{2\Delta kL}}
{|(1{+}\kappa)^2 - (1{-}\kappa)^2 e^{2i\Delta kL}|^2} \,.
\end{multline}
$G^{(c)}_x(\theta\!=\!0)$ and $G_y^{(s,\hat\bfn_\ux)}(\theta\!=\!0)$ are
plotted in Fig.~\ref{bsc::fig:3} as functions of
$E_F^*$ and $\alpha/\beta$ for different crystal orientations $\xi$.
The peaks in $G^{(c)}_x(\theta\!=\!0)$ and
$G_y^{(s,\hat\bfn_\ux)}(\theta\!=\!0)$ as a function of $E_F^*$ come from
the Fabry-Perot interference, which gives rise to resonances for
\begin{equation}
\label{eq:peaks}
\Delta k L = n\pi \qquad(n=0,1,2,\ldots) \,.
\end{equation}
Unlike the (longitudinal) charge current, the spin current is very
sensitive to the SO coupling strengths, $\alpha$ and $\beta$, and the
crystal orientation, $\xi$, as seen from the factor $\alpha\beta\cos\xi$
in Eq.~\eqref{eq:gsyn}.

\section{Angle-Averaged Conductances}

For true one-dimensional (1D) leads ($k_FW\ll 1$), where only a single
transverse mode is allowed, one has only to consider normal incidence
($\theta\!=\!0$); or at a certain fixed $\theta$\cite{Marigliano04a}.
In the opposite limit (i.e., $k_FW\to\infty$), there are many transverse
modes contributing to the transport.  In this case, we should add up all
the contributions from $\theta$ in the range $(-\pi/2,\pi/2)$.  It is
quite complicated (even though possible) to find the scattering states
for non-zero incidence angle $\theta$, and more convenient to work
numerically.  Therefore, here we just present the numerical results.

\begin{figure}
\centering%
\includegraphics*[width=85mm]{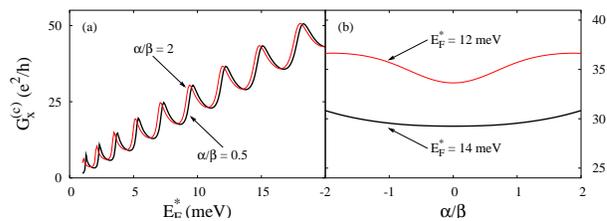}
\caption{Angle-averaged charge conductance $G^{(c)}_x$ as
  a function of (a) $E_F^*$ and (b) $\alpha/\beta$.  $\xi=0$ has been
  chosen, but $G_x^{(c)}$ is not sensitive to $\xi$.  Values of other
  parameters are indicated in the figures.}
\label{bsc::fig:4}
\end{figure}

Apparently, the longitudinal charge current has a main contribution from
the normal incidence.  Consequently, as shown in Fig.~\ref{bsc::fig:4}
the $\theta$-averaged longitudinal conductance $G_x^{(c)}$ is rather
similar to the normal incidence case $G_x^{(c)}(\theta\!=\!0)$, although the
peaks are rounded off.

\begin{figure}
\centering%
\includegraphics[width=8.5cm]{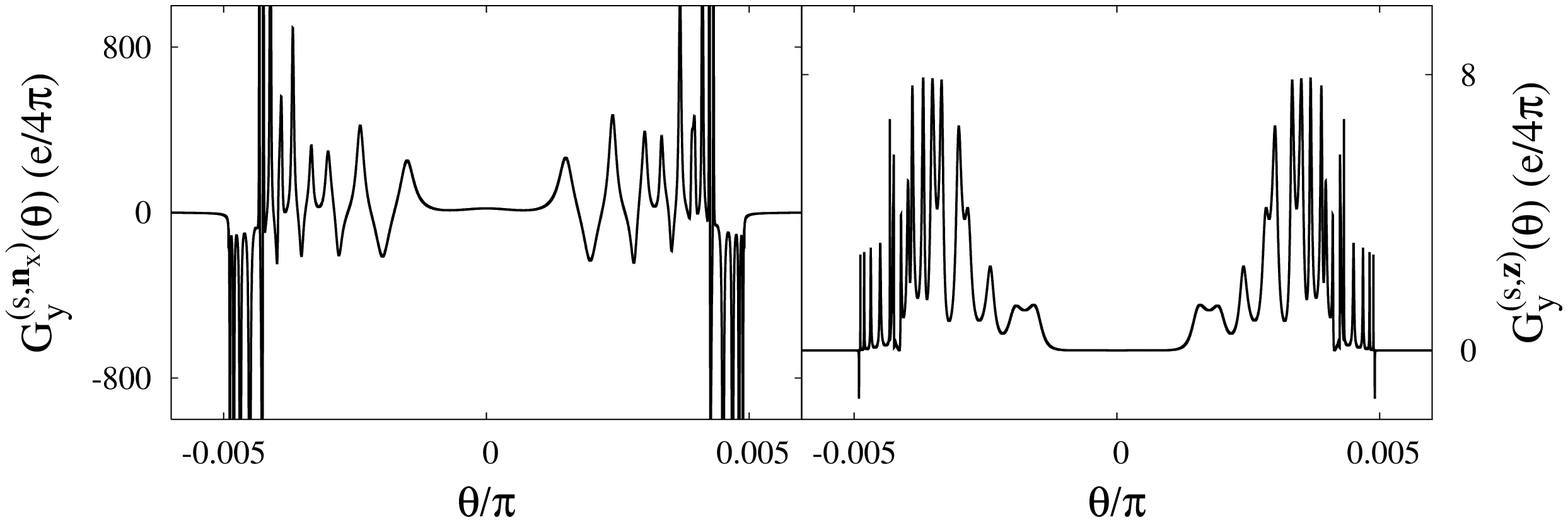}
\caption{Angle dependences of the spin conductances
  $G_y^{(s,\hat\bfn_\ux)}(\theta)$ and $G_y^{(s,\hat\bfz)}(\theta)$ for
  $E_F^*=14\,\meV$, $\alpha/\beta=0.5$, and $\xi=0$.}
\label{bsc::fig:5}
\end{figure}

\begin{figure}
\centering %
\includegraphics*[width=85mm]{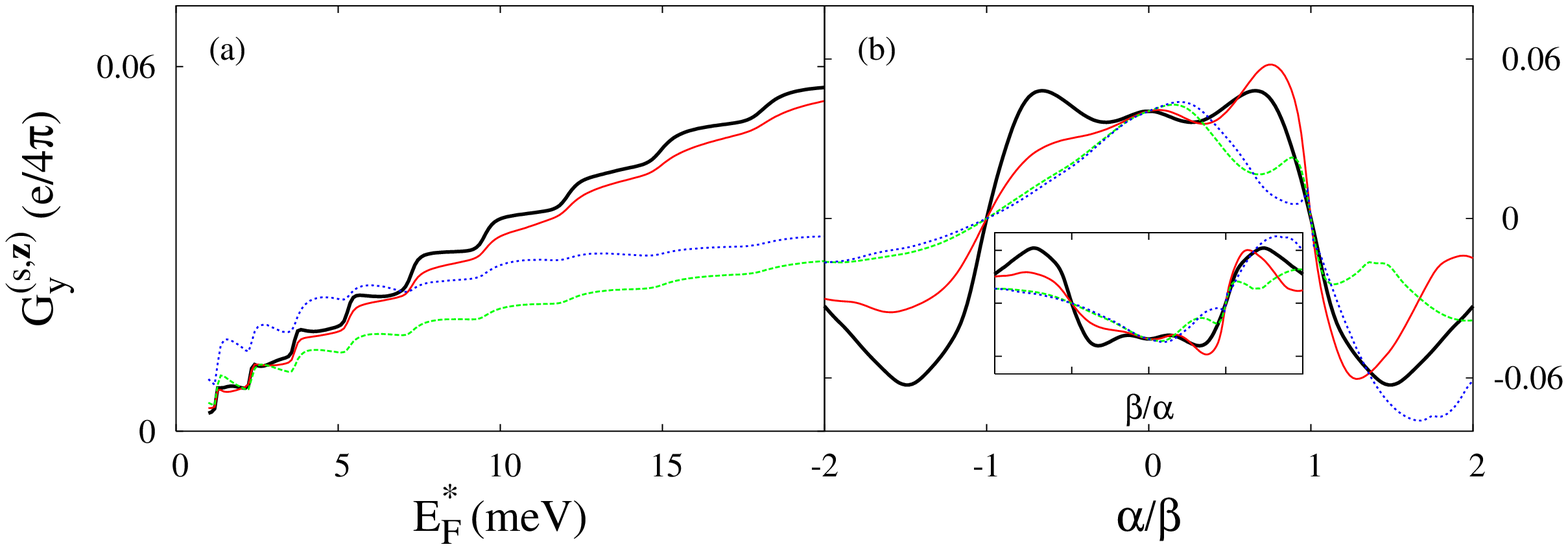}
\caption{Transverse spin conductance $G_y^{(s,\uz)}$ as a function of
  (a) $E_F^*$ for $\alpha/\beta=0.5$ and (b) $\alpha/\beta$ for
  $E_F^*=14\,\meV$. Inset: $G_y^{(s,\uz)}$ as a function of
  $\beta/\alpha$ with $\alpha=0.1\,\eV{\cdot}\angstrom$ fixed. The
  values of $\xi$ are the same as in \figref{bsc::fig:3}.}
\label{bsc::fig:6}
\end{figure}

This is not the case for the spin transport.
\Figref{bsc::fig:5} shows the $\theta$-dependence of the
spin conductances polarized in the $\hat\bfn_\ux$ and $\hat\bfz$,
respectively.  Again, the peaks correspond to the Fabry-Perot-type
resonances.  When summing up, the contributions to the
$\hat\bfn_\ux$-polarized spin current from different angles are
mostly canceled with each other, and hence the angle-averaged spin
conductance $G_y^{(s,\hat\bfn_\ux)}$ becomes very small compared
with the longitudinal charge conductance $G_x^{(c)}$.
On the other hand, the $\hat\bfz$-polarized spin current is not subject
to such cancellations, and remains rather large (still smaller than the
longitudinal charge current); see Fig.~\ref{bsc::fig:6}.
Especially for $\xi=0$, the spin conductance $G^{(s,\uz)}_y$ remains
almost constant in the region $|\alpha/\beta|<1$ and changes its sign
abruptly at $\alpha/\beta=\pm1$. This behavior is reminiscent of the
intrinsic spin Hall conductances in the previous
works.\cite{ISHE1,ISHE2} However, in our case $G^{(s,\uz)}_y$ depends on
the strength of the SO couplings, the potential barrier, crystal
orientation, and the channel length, showing no universal
characteristics.

Here we stress the differences between origins in the spin Hall
conductance of ours and the intrinsic spin Hall effect.
The intrinsic spin Hall effect is an (semi-classical) effect
driven by external electric field penetrating the (infinitely large)
system.~\cite{ISHE1,ISHE2}  In our case, the external bias voltage
merely shifts the relative chemical potentials of the ``contacts'' (or
reservoirs) attached to the metallic leads where the electrons undergo
ballistic transport and does not feel an electric field.~\cite{Datta95a}
Moreover, it has been pointed out that the spin current in the intrinsic
spin Hall effect is an equilibrium background current and is not
real.\cite{Rashba03} To the contrary, our spin currents originate from
non-equilibrium properties of the system, and are real.

We also note that when electrons are incident oblique to the junction
interface, the transverse charge current and the longitudinal spin
current do not vanish any longer.  Therefore, the angle-averaged
conductances $G^{(c)}_y$ and $G^{(s,\hat\bfn)}_x$ are finite, even though
quite small compared to $G^{(c)}_x$.
Finite $G^{(c)}_y$ in the semiconductor can be attributed to the
anisotropy introduced by the Dresselhaus effect. It distorts the group
velocity, which thus prefers one of $\pm y$ directions so that the
current has same sign as $\alpha\beta$.
Nonzero $G^{(s,\hat\bfn)}_x$ reflects the breaking of spin conservation
inside the semiconductor. For oblique incidence, there exists no
direction consistent with the boundary conditions along which spin state
is stationary (for example, the spin parallel to the direction
$\hat\bfn_\bfk$ for a given wave vector $\bfk=k_x\ux+k_y\uy$
is not stationary any longer for $\bfk'=-k'_x\ux+k_y\uy$ after
reflection from the junction interface).  This means that an electron
with any spin polarization experiences precession during transmission
through the semiconductor.

\section{Conclusion}

Ballistic spin currents with different spin polarizations through
mesoscopic metal/2DES/metal double junctions have been investigated in
the presence of spin-orbit interactions.  Using the coherent scattering
theory we showed that longitudinal and/or transverse spin currents can
flow through 2DES.  It was argued that arising from the non-equilibrium
distribution of electrons, the spin Hall currents observed are real.

\section*{Acknowledgments}
M.L.\ thanks W.~Belzig, C.~Bruder, and J.~Schliemann for
helpful discussions.  This work has been supported from the SKORE-A
program and the eSSC at Postech.  M.-S.C.\ acknowledges the support from
KIAS, where part of the work was done.

\end{document}